\newcommand{\be}{\begin{equation}}\newcommand{\ee}{\end{equation}}
\newcommand{\bea}{\begin{eqnarray}}\newcommand{\eea}{\end{eqnarray}}
\newcommand{\nn}{\nonumber}\newcommand{\p}[1]{(\ref{#1})}
\newcommand{\lb}[1]{\label{#1}}
\newcommand\s{\scriptscriptstyle}

\newcommand\q{\quad}
\newcommand\qq{\quad\quad}
\renewcommand\={\ =\ }

\newcommand\cD{{\cal D}}

\newcommand\tpa{\theta^{+\alpha}}

\newcommand\btpa{\bar{\theta}^{+\dot{\alpha}}}

\newcommand\bth{\bar{\theta}}
\newcommand\tp{\theta^+}

\newcommand\btp{\bar\theta^+}

\def\a{\alpha}
\def\da{{\dot\alpha}}
\def\b{\beta}

\def\g{\gamma}

\def\d{\delta}

\def\eps{\epsilon}

\def\ve{\varepsilon}

\def\bph{{\bar\phi}}
\def\vp{\varphi}

\def\j{\psi} 
\def\bj{{\bar\psi}}

\def\th{\theta}  
\def\bt{\bar\theta}

\def\si{\sigma}

\def\J{\Psi}
\def\bJ{\bar\Psi}
\def\L{\Lambda}

\def\pa{\partial}

\newcommand\ab{{\alpha\beta}}

\newcommand\ada{{\alpha\dot{\alpha}}}

\newcommand\pada{\partial_{\alpha\dot{\alpha}}}

\newcommand\A{{\s A}}

\newcommand\C{{\s C}}

\newcommand\sL{{\s L}}

\newcommand\W{{\s W}}
\newcommand\Z{{\s Z}}

\newcommand{\pp}{{++}}

\newcommand{\Dp}{D^{\pp}}

\newcommand{\Vp}{V^\pp}

\newcommand{\Dpa}{D^+_\alpha}

\newcommand{\bDpa}{\bar{D}^+_{\dot{\alpha}}}

\def\sfrac#1#2{{\textstyle\frac{#1}{#2}}}

\def\sha{\sfrac12}

\newcommand{\mR}{\mathbb R}
\def\e{\mbox{e}}
\def\ii{\mbox{i}}
\def\diff{\mbox{d}}

\documentclass[12pt]{article}
\usepackage{amsmath,amssymb}

\begin{document}

\begin{titlepage}
\begin{flushright}

ITP-UH-21/04\\
hep-th/0408146
\end{flushright}

\vspace{1cm}

\begin{center}

{\Large\bf
NON-ANTICOMMUTATIVE DEFORMATION \\[8pt]
OF N=(1,1) HYPERMULTIPLETS}

\vspace{1.5cm}

{\large\bf
Evgeny Ivanov$\,{}^a$, \ Olaf Lechtenfeld$\,{}^b$, \ Boris Zupnik$\,{}^a$}
\vspace{1cm}

${}^a${\it Bogoliubov Laboratory of Theoretical Physics, JINR, \\
141980, Dubna, Moscow Region, Russia}\\
{\tt eivanov, zupnik@thsun1.jinr.ru}\\[8pt]

${}^b${\it Institut f\"ur Theoretische Physik, Universit\"at Hannover, \\
30167 Hannover, Germany}\\
{\tt lechtenf@itp.uni-hannover.de}

\end{center}

\vspace{2cm}

\begin{abstract}

\noindent
We study the SO(4)$\times$SU(2) invariant and $N{=}(1,0)$
supersymmetry-preserving nil\-potent (non-anticommutative) Moyal
deformation of hypermultiplets interacting with an abelian gauge
multiplet, starting from their off-shell formulation in Euclidean
$N{=}(1,1)$ harmonic superspace. The deformed version of a neutral
or a charged hypermultiplet corresponds to the `adjoint' or the
`fundamental' representation of the deformed U(1) gauge group on
the superfields involved. The neutral hypermultiplet action is
invariant under $N{=}(2,0)$ supersymmetry and describes a deformed
$N{=}(2,2)$ gauge theory. For both the neutral and the charged
hypermultiplet we present the corresponding component actions and
explicitly give the Seiberg-Witten-type transformations to the
undeformed component fields. Mass terms and scalar potentials for 
the hypermultiplets can be generated via the Scherk-Schwarz mechanism 
and Fayet-Iliopoulos term in analogy to the undeformed case.

\end{abstract}

\end{titlepage}

\setcounter{page}{1}

\section{Introduction}

Some backgrounds in string theory are known to trigger various types of 
space-time noncommutativity in the low-energy limit. 
For instance, a constant Neveu-Schwarz 
$B$-field background in type IIB string theory implies that the relevant 
low-energy dynamics is described by a gauge theory defined on non-commutative 
flat space, with $[x^i, x^j] = \ii\theta^{ij}$, where $\theta^{ij}$ is a 
constant skew-symmetric matrix \cite{SW} (see also review \cite{Rev1}). 
Recently it was discovered that certain string backgrounds 
give rise to supersymmetric field theories living on superspaces with 
non-anticommuting Grassmann coordinates \cite{OVa}--\cite{BS}. In particular, 
a specific four-dimensional compactification of the type IIB string in the 
presence of a constant self-dual graviphoton background $F^{\alpha\beta}$ 
yields a superspace whose odd coordinates obey the Clifford algebra 
$\{\theta^\alpha,\theta^\beta\}=\alpha{}'{}^2F^{\alpha\beta}$ rather than 
the standard Grassmann algebra \cite{Se}. 
This superspace and supersymmetry realized in it must be of Euclidean 
signature since a real field strength can be self-dual 
($F^{\alpha\beta}{\neq}0,F^{\dot\alpha\dot\beta}{=}0$) only in Euclidean 
(or Kleinian) but not in Minkowski space. The classical and quantum properties
of theories defined on such nilpotently-deformed (or non-anticommutative) 
Euclidean $N{=}(\sha,\sha)$ superspace were analyzed in \cite{Se,BS} and 
in many subsequent works (see \cite{Rev2} for a review). As its most 
characteristic feature, this type of deformation breaks the original 
$N{=}(\sha,\sha)$ supersymmetry by half, i.e.~down to $N{=}(\sha,0)$, 
but preserves the important notion of chirality. The basic technical device 
of constructing the corresponding superfield theories is the Moyal-Weyl star 
product extended to Grassmann coordinates \cite{FL,KPT,FLM}. 

A natural step beyond the analysis of non-anticommutative $N{=}(\sha,\sha)$ 
superspace is the investigation of analogous nilpotent chiral deformations 
of Euclidean $N{=}(1,1)$ superfield theories. This study was initiated in 
\cite{ILZ,FS} and then continued and advanced in \cite{FILSZ}--\cite{KS}. 
Both the D-type and Q-type deformations were considered, with either spinor 
covariant derivatives $D^i_\alpha$ or supersymmetry generators $Q^i_\a$ as 
the building blocks of the bi-differential Poisson operator defining the 
relevant star products \cite{FL,KPT,FLM}.

The Q-deformations generically break the $N{=}(1,1)$ supersymmetry by half,
i.e.~down to $N{=}(1,0)$, but preserve both chirality and anti-chirality.
The simplest $N{=}(1,1)$ Q-deformation is the singlet one (`QS-deformation'), 
based on the Poisson operator
\be
P_s=-I\overleftarrow{Q}{}^i_\a\overrightarrow{Q}{}_i^\a \ ,
\qquad\text{with}\quad (P_s)^5=0 \ ,
\ee
where $I$ is a real parameter. While breaking half the supersymmetry, 
it preserves the internal SU(2)$_R\times $Spin(4) symmetry,
something impossible in the $N{=}(\sha, \sha)$ case. Furthermore, the
QS-deformation as well can be given a stringy interpretation \cite{FILSZ} 
along the line of \cite{Se,OVa,BGN,BFPL}. Namely, such a non-anticommutative 
$N{=}(1,1)$ superspace naturally arises for the $N{=}4$ superstring coupled 
to a complex axion background. The Q-deformations and their QS-subclass also 
preserve Grassmann harmonic analyticity \cite{ILZ,FS} which is the fundamental
notion in theories with manifest extended supersymmetry \cite{GIK1,GIOS}.
In view of these motivations, a complete understanding of the geometric 
and quantum structure of Q-deformed $N{=}(1,1)$ theories 
(and of their higher-$N$ counterparts) is as important as the exploration of 
the analogous properties of non-anticommuting $N{=}(\sha,\sha)$ theories.  
    
Until now, the main focus was on the Q-deformation of $N{=}(1,1)$ pure gauge
theories. In \cite{FILSZ}, the detailed superfield and component structure
of the QS-deformed $N{=}(1,1)$ U(1) and U($n$) gauge theories was
explored.\footnote{The QS-deformed U(1) theory was independently
considered in \cite{AI}.} In particular, an analog of the Seiberg-Witten
(SW) map to quantities with undeformed gauge and supersymmetry
transformation laws, both for the component fields and for the off-shell
superfield strengths, was explicitly worked out.

An important (and not yet well-studied) class of $N{=}(1,1)$ theories are 
those including matter hypermultiplets interacting with themselves and with 
gauge multiplets. These theories are analogs of the coupled systems of chiral
and gauge superfields of the $N{=}(\sha,\sha)$ case \cite{Se}. The first type
of theories, i.e.~self-interacting hypermultiplets, yields, in the bosonic 
sector, Euclidean versions of hyper-K\"ahler sigma models. The second type, 
i.e.~hypermultiplets coupled to $N{=}(1,1)$ gauge multiplets, could be of 
relevance from the phenomenological point of view. The system of a gauge 
superfield minimally coupled to a hypermultiplet in the adjoint representation
of the gauge group provides an off-shell $N{=}(1,1)$ superfield formulation of
$N{=}(2,2)$ supersymmetric gauge theory which is the Euclidean 
analog of the renowned $N{=}4$ super Yang-Mills theory. With these motivations
in mind, it is of obvious interest to elaborate on the structure of
Q-deformed $N{=}(1,1)$ hypermultiplet theories.
Note that, in contrast, the singlet D-deformation (which is
the unique D-type deformation preserving Grassmann harmonic analyticity
\cite{ILZ,FS}) does not at all affect $N{=}(1,1)$ analytic superfield
Lagrangians, including those for hypermultiplets. So only the
Q-deformations are of interest in the hypermultiplet context which is the
subject of this paper.

In \cite{ILZ} we gave general recipes for constructing Q-deformations of
superfield hypermultiplet actions but did not work out specific interesting 
examples of such models.
In the present paper we consider, both in the superfield and component-field 
formulations, the QS-deformation of two simple actions with hypermultiplet and
U(1) gauge superfields.\footnote{A toy model of QS-deformed self-interacting
hypermultiplets, with the interaction vanishing in the undeformed limit, was
recently considered in \cite{IZ}.}

Our first example, considered in Section~4, is the coupled system of an
$N{=}(1,1)$ U(1) gauge superfield and a neutral hypermultiplet. Before turning
on the QS-deformation, the action of this system is simply the sum
of free actions for the gauge and hypermultiplet superfields, with one extra
hidden on-shell $N{=}(1,1)$ supersymmetry extending the manifest $N{=}(1,1)$
supersymmetry to $N{=}(2,2)$. Once the deformation is turned on, the
hypermultiplet starts to transform in a non-trivial way under the deformed
U(1) gauge group: this transformation coincides with the `non-abelian' part
of the gauge superfield transformation. We pass to the physical component
fields and find the corresponding SW-transformation. In terms of the
undeformed fields the action radically simplifies.
Yet, there still remains a new non-trivial interaction
between the fermionic fields of the hypermultiplet and the gauge field
(alongside with the known interaction between the gauge field and one of
the scalar fields of the gauge multiplet \cite{FILSZ,AI}).
Besides the manifest unbroken $N{=}(1,0)$ supersymmetry, the
resulting action proves to possess one more hidden $N{=}(1,0)$ supersymmetry
with on-shell closure. Thus we are left with an unbroken $N{=}(2,0)$
supersymmetry in the QS-deformed $N{=}(2,2)$ U(1) gauge theory.\footnote{
Nilpotent Q-deformations of the on-shell superfield constraints
of the Euclidean $N{=}(2,2)$ gauge theory were studied in \cite{SWo}.}

Section~5 is devoted to the QS-deformation of a charged hypermultiplet
in minimal interaction with a U(1) gauge multiplet. In the undeformed case, 
this interacting system possesses no extra supersymmetry besides the manifest
$N{=}(1,1)$ one. The QS-deformation breaks the latter down to $N{=}(1,0)\,$.  
We analyze the component action of this deformed $N{=}(1,1)$ electrodynamics 
and the corresponding scalar field potential.
The SW-transformation to undeformed fields exists in this case too. 

An interesting notable feature of the two models considered is that
there are only two inequivalent ways to implement the deformed U(1) gauge
transformations on the hypermultiplet superfields.
In Section~4 we deal with the `adjoint' representation, while the deformed
charged representations of Section~5 are equivalent to the `fundamental'
representation. One can treat a pair of mutually conjugate analytic
hypermultiplet superfields as a doublet of an additional rigid
`Pauli-G\"ursey' SU(2)$_{PG}$ group \cite{GIOS}. The `adjoint'
U(1) transformation commutes with SU(2)$_{PG}$, while the `fundamental'
U(1) transformation commutes only with U(1)$_{PG}\subset$ SU(2)$_{PG}$
(just as in the undeformed case). Respectively, the first and the second
model possess SU(2)$_{PG}$ and U(1)$_{PG}$ global symmetries in parallel
with their SU(2)$_R$ and Spin(4) automorphisms.

%\vfill\eject

\setcounter{equation}0
\section{QS-deformation of N=(1,1) U(1) gauge theory}

We start with a short overview of the Q-deformed Euclidean harmonic
$N{=}(1,1)$ superspace and $N{=}(1,1)$ U(1) gauge theory following refs.
\cite{ILZ} and \cite{FILSZ}.

The basic concepts of the $N{=}(1,1), D{=}4$ Euclidean harmonic superspace,
which is an extension of the standard $N{=}(1,1)$ superspace by the
SU(2)$_R$/U(1)$_R$ harmonics $u^\pm_i$, are collected in \cite{ILZ,FS,FILSZ}
(see also \cite{GIOS}).
The standard (central) cordinates of the $N{=}(1,1)$ harmonic superspace
are $(Z,u^\pm_i)=(x^m,\th^\a_k,\bth^{\da k},u^\pm_i)\,$.
We shall also use the  chiral coordinates
\be
Z_\sL=(x^m_\sL, \th^\a_k, \bt^{\da k})\, , \q x^m_\sL = x^m+
\ii(\sigma^m)_\ada\th^\a_k\bt^{\da k}\,,
\ee
the chiral-analytic coordinates
\be
 Z_\C=(x^m_\sL,\theta^{\pm\alpha}\, ,
 \bar\theta^{\pm\da}),
\ee
and the analytic coordinates
\bea
Z_\A=(x^m_\A, \th^{\pm\a}, \bt^{\pm\da})\,, \q
x^m_\A\=
x^m_\sL-2\ii(\sigma^m)_\ada\theta^{-\alpha}\btpa \,,
\lb{Acoor}
\eea
where $\th^{\pm\a}=u^\pm_k\th^{\a k}$ and $\bt^{\pm\da}=u^\pm_k\bt^{\da k}$.
It should be stressed that all these coordinates are (pseudo)real with
respect to the basic conjugation $\;\widetilde{\;\;} \;$ \cite{ILZ}. For instance, the 
condition of
reality can be consistently imposed on the Euclidean chiral superfields.
An important invariant pseudoreal subspace of the harmonic superspace
$(Z_\A, u^\pm_i)$ is the analytic harmonic superspace
$$
(\zeta, u^\pm_i) \,,
$$
where
$$
\zeta=(x^m_\A,\tpa,\btpa)\,.
$$

The supersymmetry-preserving spinor derivatives $D^\pm_\a, \bar{D}^\pm_\da$
and the harmonic derivatives $D^{\pm\pm}$ in different coordinate bases are given in 
\cite{ILZ,FILSZ}.
An analytic superfield satisfies the conditions
\be
(\Dpa, \bDpa) \Phi(Z, u^\pm_i) \= 0 \quad \leftrightarrow \quad \Phi =
\Phi_\A(\zeta, u^\pm_i)\,.
\ee

In what follows it will be convenient to use harmonic projections of the
supersymmetry generators
\be
Q^k_\alpha=u^{+k}Q^-_\alpha-u^{-k}Q^+_\alpha,\q
\bar Q_{\da k}=u^+_k\bar Q^- -u^-_k\bar Q^+.
\ee
For instance, in the analytic coordinates
\be
Q^+_\a=\partial_{-\a}-2\ii\btpa\pada,\q Q^-_\a=-\partial_{+\a}\,, \lb{Qgen}
\ee
where $\partial_{\pm\alpha}=\partial/\partial\th^{\pm\alpha}$ and
$\pada=(\sigma_m)_\ada\pa_m$.

In this paper we shall deal with the Q-singlet (or QS-) deformation
which is associated with the SO(4)$\times$SU(2) invariant Poisson bracket
\be
AP_sB=-I(-1)^{p(A)}Q^k_\a A Q^\a_k B=-I(-1)^{p(A)}\,(Q^+_{\a} A Q^{-\a}
B+Q^{-\a} A Q^+_{\a} B).\lb{Psingl}
\ee
The noncommutative QS-star product of two analytic superfields has the
following simple form:
\bea
&&\L \star\Phi=\L\Phi+\L P_s\Phi+\sha\L P^2_s\Phi\,. \lb{ProdAnal}
\eea
Since $\partial^\alpha_-A =\partial^\alpha_-B = 0$ in the analytic basis,
we can omit $\partial_{-\alpha}$ in $Q^+_\alpha$ in this basis. In particular,
\be
A P_s B = 2\ii I(-1)^{p(A)}\btpa(\pa^\a_+ A \pada B-
\pada A \pa^\a_+ B)\,. \lb{APBanal}
\ee
In the $\star$-commutator of two bosonic analytic
superfields $[A, B]_\star$ with $[A, B]=0$ only the term $\sim P_s$ survives:
\be
[A, B]_\star = 2AP_sB\,, \lb{CommStar}
\ee
while the $\star$-anticommutator reads
\be
\{A, B\}_\star = 2AB + A P^2_s B .
\lb{AntcommStar}
\ee
We shall make use of these general formulas in Sections 4 and 5.

The basic superfield of the $N{=}(1,1)$ gauge theory is the
analytic gauge potential $\Vp$. The QS-deformed gauge transformation of the
U(1) potential $\Vp$ in the analytic basis reads
\bea
&& \delta_\L\Vp\= \Dp\L+ [\Vp, \L]_\star = \Dp\L+2\Vp P_s\L \nn \\
&& =\,\Dp\L+4\ii I\btpa(\pa^\a_+\Vp\pada\L-
\pada\Vp\pa^\a_+\L)\,. \lb{U1V}
\eea
Here $\L$ is an  analytic gauge parameter and $\tilde\L=-\L$. The gauge
potential in the Wess-Zumino gauge has the following $\tpa, \btpa$-expansion
\bea
&&\Vp_{\W\Z} =(\tp)^2\bar\phi+(\btp)^2\phi+2\tp\si_m\btp A_m+4(\tp)^2
\btp_\da u^-_k\bJ^{\da k}\nn\\
&&+\,4(\btp)^2\tpa u^-_k\J^k_\a+3(\tp)^2(\btp)^2u^-_ku^-_l\cD^{kl}\,,\lb{WZ}
\eea
with all the component fields being functions of~$x^m_\A$.
The residual U(1) gauge transformations with the parameter
$\L_r=ia(x_\A)$ act on the component fields in \p{WZ} as
\bea
&&\d_r \bph=0,\q\d_r \phi=-8IA_m\pa_m\,a\,,\q\d_r A_m
=(1+4I\bph)\pa_m\,a\,,\nn\\
&&\d_r\J^k_\a=-4I\bJ^{\da k}\pada a\,\q \d_r\bJ^k_\da=0\,,\q \d_r\cD^{kl}=0\,.
\lb{ResComp}
\eea

The deformed $N=(1,0)$ supersymmetry transformations of the U(1) gauge
multiplet component fields defined in \p{WZ} are as follows \cite{FILSZ}:
\bea
&&\delta_\epsilon\phi=2\epsilon^{\a k}\Psi_{\a k}\,,
\q\delta_\epsilon\bph=0\,,\q
\delta_\epsilon A_m= \epsilon^{\a k}(\sigma_m)_\ada\bar\Psi_k^\da\,,\nn\\
&&\delta_\epsilon\Psi^k_\a=
-\epsilon_{\a l}\cD^{kl}+\sfrac12(1+4I\bph)(\si_{mn}\eps^k)_\a
F_{mn}-4\ii I\epsilon^k_\a A_m\pa_m\bph\,,\nn\\
&&\delta_\epsilon\bar\Psi^k_\da=-\ii\epsilon^{\a k}(1+4I\bph)\pada\bph\,,\nn\\
&&\delta_\epsilon \cD^{kl}=
\ii\pa_m[(\epsilon^k\si_m\bar\Psi^l+\epsilon^l\si_m\bar\Psi^k)
(1+4I\bph)] \,, \lb{defsusy}
\eea
where $F_{mn}=\pa_mA_n-\pa_nA_m\,$. They are produced by the transformation
of $\Vp_{\W\Z}$ which is a sum of the standard supertranslation piece and
the compensating gauge transformation with the parameter
\bea
&\Lambda_\eps=2 (\epsilon^-\theta^+) \bph-2\btp_\da\eps^-_\a( A^\ada
+2\tpa\bJ^{-\da})+2(\btp)^2[(\eps^-\Psi^-)+(\eps^-\th^+)\cD^{--}]\,.&
\lb{WZpar}
\eea

The Seiberg-Witten (SW) transformation to the undeformed U(1) gauge
supermultiplet $\vp,\bph,a_m,\j^\a_k,\bj^\da_k, d^{kl}$ with the standard
gauge and $N{=}(1,0)$ supersymmetry transformation properties (corresponding
to the choice $I{=}0$ in \p{ResComp}, \p{defsusy}) is defined as the
following set of relations
\bea
&&\phi=(1+4I\bph)^2\vp-4I(1+4I\bph)^{-1}[A_m^2+4I^2(\pa_m\bph)^2]\,,\nn\\
&&A_m=(1+4I\bph)a_m\,,\q\bJ^k_\da=(1+4I\bph)\bj^k_\da\,,\nn\\
&&\J^k_\a=(1+4I\bph)^2\j^k_\a-4I(1+4I\bph)a_\ada\bj^{\da k}\,,\nn\\
&&\cD^{kl}=(1+4I\bph)^2d^{kl}-8I(1+4I\bph)\bj^k_\da\bj^{\da l}\,.
\lb{SWtran}
\eea
Using them, one can express the component Lagrangian of the deformed U(1)
gauge theory through the standard undeformed free $N{=}(1,1)$ gauge theory
Lagrangian \cite{FILSZ,AI}
\be
L_g=(1+4I\bph)^2[-\sha\vp\,\Box\bph+\sfrac14 f^2_{mn}+\sfrac18\ve_{mnrs}
f_{mn}f_{rs}-\ii\j^\a_k\pada\bj^{\da k}+\sfrac14(d^{kl})^2] \,,\lb{LAGRgauge}
\ee
where $f_{mn}=\pa_ma_n-\pa_na_m$.  After redefining the involved fields as
\be
\hat{\varphi} = (1+4I\bph)^2\vp\,, \;\;\hat{\psi}{\,}^\alpha_k =
(1+4I\bph)^2\j^\a_k\,,\;\;\hat{d}{\,}^{kl} = (1+4I\bph)d^{kl}\,,
\ee
the Lagrangian \p{LAGRgauge} acquires the form in which it differs from the free Lagrangian 
only by
a simple interaction term:
\be
L_g= -\sha\hat{\vp}\,\Box\bph -\ii\hat{\j}{\,}^\a_k\pada\bj^{\da k}
+\sfrac14 (1+4I\bph)^2\left(f^2_{mn}+\sfrac12\ve_{mnrs}
f_{mn}f_{rs}\right) +\sfrac14(\hat{d}{\,}^{kl})^2\,. \lb{LAGRgauge1}
\ee
Note that the fields $\bph,  \hat{\psi}{\,}^\alpha_k, \bj^\da_k$ and
$\hat{d}{\,}^{kl}$ satisfy the free equations.

\setcounter{equation}0
\section{Representations of the QS-deformed gauge group}

The free $q^+$  hypermultiplet action of the ordinary harmonic theory
\cite{GIOS} is not deformed in the non-anticommutative superspace
\cite{ILZ,FS}:
\be
S_0(q^+)\=-\int\!\diff u\,\diff\zeta^{-4}\ \tilde{q}{\,}^+ \Dp q^+=
\sfrac12\int\!\diff u\,\diff\zeta^{-4}\ q^+_a \Dp q^{+a}\,.
 \lb{freeAc}
\ee
Here $\diff\zeta^{-4}=\diff^4x_\A (D^-)^4$ and the additional SU(2)$_{PG}$
indices $a, b=1, 2$ are introduced, $q^{+a}=\ve^{ab}q^+_b=(\tilde{q}^+, q^+)=
\widetilde{q^+_a}$.
After passing to the component fields, integrating over $\theta^+_\a,
\bar\theta^+_\da$ and eliminating the infinite tower of the auxiliary fields
by their algebraic equations of motion, the superfield free action \p{freeAc}
yields the free action for (4+8) physical fields of the hypermultiplet, viz.
scalars $f^{ak}(x_\A)$ and fermions $\rho^a_\a(x_\A)$ and $\chi^a_\da(x_\A)$.
We do not quote it here; it can be easily reproduced as the $I\rightarrow 0$
limit of the QS-deformed actions to be given below.

Let us discuss how the QS-deformed U(1) gauge transformations can be
implemented on the superfield $q^{+a}$. Obviously, these transformations
should have the same Lie bracket structure as the the QS-deformed U(1) gauge
transformations \p{U1V} of $V^{++}$. It is easy to see that the Lie bracket
of two such transformations has again the form \p{U1V} with $\L_{br}=
[\L_2,\L_1]_\star$. Then a simple analysis shows that only two non-equivalent
realizations of the deformed U(1) gauge group with such Lie bracket structure
are possible for the hypermultiplet doublet $q^{+a}$ (besides the trivial
realization $\delta_\L q^{+a} = 0$)
\bea
&&1. \q\d_\L q^{+a}=[q^{+a}, \L]_\star\,,\lb{adj}\\
&&2. \q \d_\L q^{+a}= \sha[q^{+a}, \L]_\star+\sha(\tau_3)^a_b\{q^{+b},
\L\}_\star \,, \lb{fund}
\eea
where $\tau_3$ is the diagonal Pauli matrix. These representations of the
deformed U(1) group can be naturally called `adjoint' and `fundamental',
respectively. Indeed, \p{adj} has the same form as the non-abelian part of
the deformed U(1) transformation \p{U1V} of the gauge superfield,
while \p{fund} can be equivalently rewritten as
\be
\d_\L q^+=-\L\star q^+\,, \q \d_\L \tilde{q}^+ = \tilde{q}^+ \star \L \,.
\lb{fund2}
\ee

An interesting peculiarity of the QS-deformed realization \p{fund}, \p{fund2}
of the U(1) gauge group is that the U(1) charge of $q^+$ and $\tilde{q}^+$
in it is fixed up to a finite SU(2)$_{PG}$ rotation of this pair. In the
commutative limit we are left with the U(1) charges $\mp 1$  for $q^+$ and
$\tilde{q}^+$. On the contrary, in the undeformed case one can ascribe to
$q^+$ an arbitrary real charge $e$ corresponding to the U(1) gauge
transformation $\delta_\L q^+ = e\,\L q^+\,, \;\delta_\L \tilde{q}^+
= - e\,\L \tilde{q}^+\,$. This difference is related to the fact that
QS-deformed U(1) transformations have a non-trivial closure on off-shell
superfields (before passing to the WZ gauge \footnote{The residual U(1)
gauge group of the WZ gauge is abelian despite the fact that the full
QS-deformed superfield U(1) group has a non-trivial Lie bracket structure.})
and this closure should be the same for all involved superfields, i.e. for
$\Vp$ and $q^{+a}$. We also note that \p{fund} is a particular case of
the more general two-parameter family of realizations with the same Lie
bracket structure
\bea
&&\d_\L q^+=-\sha(\cos\mu+1)\L\star q^+ -\sha\e^{i\a}\sin\mu\,\L\star
\tilde{q}^+\nn\\
&&+\sha(1-\cos\mu)q^+\star\L-\sha\e^{i\a}\sin\mu\,\tilde{q}^+\star\L \qquad
(\mbox{and c.c.})\,,\lb{fundGen}
\eea
where $\a$ and $\mu$ are arbitrary real parameters. Eqs. \p{fund}, \p{fund2}
correspond
to the special choice $\mu=0\,$. However, it is easy to see that \p{fundGen}
can be generated from \p{fund}, \p{fund2} by a finite SU(2)$_{PG}$/U(1)$_{PG}$
rotation of the pair $(\tilde{q}^+, q^+\,)$. So all such realizations
are equivalent and, without loss of generality, we can choose \p{fund},
\p{fund2} to deal with. The `adjoint' realization \p{adj} clearly
commutes with the whole SU(2)$_{PG}$, while \p{fund} (\p{fund2}) commutes
only with U(1)$_{PG}\subset$ SU(2)$_{PG}$. So the hypermultiplet theories
associated with these two different realizations of the QS-deformed U(1)
gauge group respect SU(2)$_{PG}$ and U(1)$_{PG}$ as global internal
symmetries. In fact, the second realization can be regarded just as the
QS-deformed gauging of this latter U(1).

One can wonder whether the charged hypermultiplets with an arbitrary value of
the charge $e$ can be properly QS-deformed. The answer is affirmative, but
somewhat surprising. For this one should modify the QS-deformed U(1) gauge
transformation law \p{U1V} in the following way
\bea
&& \delta_\L\Vp\= \Dp\L+ e\,[\Vp, \L]_\star \lb{U1Vmod}
\eea
and the hypermultiplet transformation laws \p{adj} and \p{fund} as
\bea
&&1. \q\d_\L q^{+a}= e\,[q^{+a}, \L]_\star\,,\lb{adjmod}\\
&&2. \q \d_\L q^{+a}= \frac{1}{2}\,e\,[q^{+a}, \L]_\star+ \frac{1}{2}\,e\,
(\tau_3)^a_b\{q^{+b}, \L\}_\star\,.
\lb{fundmod}
\eea
Eq. \p{fundmod} can be rewritten as
\be
\d_\L q^+=-e\,\L\star q^+\,, \q \d_\L \tilde{q}^+ = e\,\tilde{q}^+ \star \L
\lb{fund3}
\ee
and it goes into the standard U(1) transformation law of the hypermultiplet
of the charge $|e|$ in the commutative limit. Any gauge transformation from
the one-parameter family \p{U1Vmod} in this limit goes into the standard
abelian transformation of $\Vp$. The modified transformation laws have the
same Lie bracket structure for all superfields, now with the bracket
parameter $e[\L_2, \L_1]_\star$. For the deformed U(1) gauge system, as well
as in the coupled system of the U(1) gauge superfield and `adjoint'
hypermultiplet, this modification is in fact unessential: the parameter $e$
can be removed from the superfield transformation laws and the relevant
invariant actions by rescaling the deformation parameter as $eI
\rightarrow I{}'$. In the system of $\Vp$ and the `fundamental'
hypermultiplet the charge $e$ cannot be removed since it is present there
already in the undeformed limit. Thus  $e$ can be treated as an additional
deformation parameter of the analytical superfield gauge group, and  gauge
transformations for different $e$ are not equivalent. Note that both in the
undeformed and deformed cases one can introduce the gauge U(1) group in such
a way that it will commute with full SU(2)$_{PG}$. This can be achieved by
introducing {\it two} independent  hypermultiplets, which
amounts to making $q^{+a}$ complex, $\widetilde{q^+_a}\neq q^{+a}$. Then
one can gauge U(1) which multiplies the whole $q^{+a}$ by phase and so
commutes with SU(2)$_{PG}$ acting on the doublet indices. The corresponding
analog of the QS-deformed transformation rules \p{fund3} is
\be
\d_\L q^{+ a}=-e\,\L\star q^{+ a}\,, \q \d_\L \tilde{q}^+_a = e\,
\tilde{q}^+_a \star\L \,. \lb{fund4}
\ee
In this paper we concentrate mainly on the simplest case
of one hypermultiplet.

The QS-deformed superfield hypermultiplet actions corresponding to the two
realizations of U(1) gauge group presented above (for arbitrary $e$) are
constructed according to the general
rule of ref. \cite{ILZ}, viz. via the replacement
\be
D^{++}q^{+\,a} \;\Rightarrow \; \nabla^{++}q^{+\,a} \lb{Repl}
\ee
in the action \p{freeAc}, where, for two options \p{adjmod}, \p{fundmod},
\bea
&& 1. \q \nabla^{++}q^{+\,a} = D^{++}q^{+\,a} + e\,[\Vp, q^{+a}]_\star =
D^{++}q^{+\,a} + 2e\, \Vp P_s q^{+\,a}\,,\lb{adjCov} \\
&& 2. \q \nabla^{++}q^{+\,a} = D^{++}q^{+\,a} + \frac{e}{2}\,
[\Vp,q^{+a}]_\star-\frac{e}{2}\,(\tau_3)^a_b\{\Vp,q^{+b}\}_\star\,.
\lb{fundCov}
\eea
The full QS-deformed action in both cases is the sum of the superfield action
of the QS-deformed U(1) gauge multiplet given in \cite{FILSZ,AI} and the
gauge invariant hypermultiplet action
\be
S(V,q) = \sfrac12\int\!\diff u\,\diff\zeta^{-4}\ q^+_a\, \nabla^{++} q^{+a}
\,. \lb{GaugeAc}
\ee

Further we shall consider the component structure of both types of the
gauge multiplet-hypermultiplet action.

\setcounter{equation}0
\section{Deformed neutral hypermultiplet}

In the explicit form,  the invariant action \p{GaugeAc} for this case reads
\bea
S_n(V,q)=\sha \int \diff u \diff \zeta^{-4} q^+_a\,\left[D^{++}q^{+a} +
4\ii I\,\btpa(\pa^\a_+\Vp\pada q^{+a}-\pada\Vp\pa^\a_+
q^{+a})\right]\,,\lb{GaugeAc1}
\eea
where we made use of eqs. \p{CommStar}, \p{APBanal} and \p{adjCov} and put
$e=1$ since the dependence on $e$ in the considered case can be absorbed
into a redefinition of $I$. To obtain the component action, we should
substitute into \p{GaugeAc1} the WZ form of $\Vp$, eq. \p{WZ}, and the
$\theta$-expansion of $q^{+a}$
\bea
&&q^{+a}= f^{+a} + \tpa\pi^a_\a + \bar\theta^+_\da\kappa^{\da a}+
\tp\si_m\btp r^{-a}_m + (\theta^+)^2 g^{-a} + (\bar\theta^+)^2 h^{-a} \nn\\
&&+(\btp)^2 \tpa \Sigma_\a^{-- a} + (\tp)^2\btpa \bar{\Sigma}^{-- a}_\da
+ (\tp)^2(\btp)^2 \omega^{-3\,a}\,,\lb{qexpan}
\eea
where all component fields are functions of $x_\A$ and $u$. Then we should
integrate in \p{GaugeAc1} over $\theta^+, \bar\theta^+$, and eliminate the
infinite number of the auxiliary fields contained in \p{qexpan} using the
appropriate non-dynamical equations of motion. It is convenient to extract
the latter directly from the superfield equation of motion
\be
\Dp q^{+a}+4\ii I\btpa(\pa^\a_+\Vp\pada q^{+a}-\pada\Vp\pa^\a_+q^{+a})=0\,.
\lb{1qeq}
\ee
Skipping details, we find the following solution for the components in
\p{qexpan} in terms of the remaining 4 physical bosonic fields $f^{ia}(x_\A)$
and 8 physical fermions $\rho^a_\a(x_\A)$, $\chi^{\da a}(x_\A)$
\bea
&& f^{+a} = f^{ak}u^+_k\,,\q\pi^a_\a = \rho^a_\a\,,\q
\kappa^{\da a} = \chi^{\da a}\,, \q r^{-a}_m  = r^{ak}_m u^-_k\,, \q g^{-a}
 = 0\,,\nn \\
&&h^{-a} = h^{ak}u^-_k\,, \q \Sigma_\a^{-- a} = \Sigma_\a^{kl\, a}u^-_ku^-_l\,,
\q\bar\Sigma_\da^{-- a} = 0\,, \q\omega^{-3\,a} =0\,, \nn \\
&& r_m^{ak}=2\ii(1+4I\bph)\pa_m f^{ak}\,,\q h^{ak}=-8\ii\,IA_m\pa_mf^{ak}\,,
\nn \\
&& \Sigma_\a^{kl\,a}=-4\ii\,I(\bJ^{\da k}
\pada f^{al}+\bJ^{\da l}\pada f^{ak})\,. \lb{Aux}
\eea
Actually, for deducing the component results below it is of no need to know
the expilcit form of the solution for $h^{-a}$ and $\Sigma^{--a}$; we have
presented it for completeness.

After substituting \p{Aux} into the original action and performing there integration
over harmonics, the final Lagrangian in the $x$-space is written in terms
of the physical fields as
\bea
&&L_1=\sfrac12(1+4I\bph)^2\,\partial_m f^{ak}\partial_m f_{ak}+
\sfrac{1}{2}\ii(1+4I\bph)\rho^{\a a}\pada\chi^\da_a
+4\ii I\bJ^\da_k\rho^\a_a\pada f^{ak}\nn \\
&&+2\ii I\rho^{\a a}A_m\pa_m\rho_{\a a}
+ \ii I\rho^{\b a}\rho^\a_a\partial_{(\alpha\dot\alpha} A^\da_{\b)}\,.\lb{L1}
\eea
Note that only the components $\bph, A_m$ and $\bJ^\da_k$ of the gauge
multiplet interact with the hypermultiplet fields in this model. We shall
see soon that in the sum of $L_1$ and the U(1) gauge multiplet Lagrangian
$L_g$, eq.\p{LAGRgauge}, most of the interaction terms can be removed by
means of the proper redefinition of fields.

Let us discuss symmetries of \p{L1}. The adjoint hypermultiplet gauge
transformation \p{adj} in the unfolded form reads
\be
\d_\L q^{+a}=4\ii I\btp_\da(\pa^\ada\L\pa_{+\a}q^{+a}-
\pa_{+\a}\L\pa^\ada q^{+a})\,,\lb{qcomm}
\ee
while the unbroken $N{=}(1,0)$ supersymmetry transformation is simply
\be
\delta_\eps q^{+\,a} = \left(\epsilon^{-\alpha}Q^+_\alpha - \epsilon^{+\alpha}Q^-_\alpha
\right)q^{+\,a} = \left(\eps^{+\a}\pa_{+\a} -2i\eps^{-\a}\btpa \pa_\ada
\right)q^{+\,a}\,,
\ee
where $\eps^{\pm \a} = \eps^{ai}u^\pm_i$. In the WZ-gauge for the U(1)
potential \p{WZ} the corresponding residual gauge and $N{=}(1,0)$
supersymmetry transformations of the hypermultiplet are
\bea
&&\d_r q^{+a}=-4I\btp_\da\pa^\ada a(x_\A)\pa_{+\a}q^{+a},\\
&&\d_\eps q^{+a}=(\eps^{+\a}\pa_{+\a} +2i\btp_\da\eps^-_\a\pa^\ada)q^{+a}+
4\ii I\btp_\da(\pa_{+\a}\L_\eps\pa^\ada q^{+a}-\pa^\ada\L_\eps\pa_{+\a}q^{+a})
\,,
\eea
where the field-dependent compensating gauge parameter  $\L_\eps$ was defined in \p{WZpar}.
After making use of eqs. \p{Aux} for the component fields in these formulas
we obtain the residual gauge transformations of the physical fields of
$q^{+a}$ as
\bea
\d_r f^{ak}=0,\q \d_r\rho^a_\a=0,\q\d_r\chi^{\da a}=-4I\pa^\ada a\,\rho^a_\a
\eea
and the corresponding $N=(1,0)$ transformations as
\bea
\d_\eps f^{ak}=\eps^{\a k}\rho^a_\a,\q\d_\eps\rho^a_\a=0,\q
\d_\eps\chi^a_\da=2\ii\eps^{\a k}(1+4I\bph)\pada f^a_k\,.\lb{10defq}
\eea

It is straightforward to construct an analog of the SW-type transform for
the physical fields of the deformed hypermultiplet
\bea
&&f^{ak}_0=(1+4I\bph)f^{ak},\qq \rho^{\a a}_0=(1+4I\bph)
\rho^{\a a},\nn\\
&&\chi^{\da a}_0=\chi^{\da a}+4I(1+4I\bph)^{-1}A^\ada\rho^a_\a-
8I(1+4I\bph)^{-1}\bJ^{\da k}f^a_k\,.\lb{SWadj}
\eea
The redefined hypermultiplet fields $f^{ak}_0, \rho^{\a a}_0, \chi^{\da a}_0$
possess the standard undeformed transformation properties: they are neutral
with respect to the gauge group U(1), i.e. their $a(x)$ variations are equal
to zero, and their $N=(1,0)$ SUSY transformations look as the $I=0$ case of
\p{10defq}.

Let us pass to the `undeformed' component fields in the Lagrangian \p{L1}.
The straightforward computation shows that, up to a total derivative, it
acquires the following simple form
\bea
&&L_1=\sfrac12\,\partial_m f_0^{ak}\partial_m f_{0\,ak}+
\sfrac{1}{2}\ii \rho^{\a a}_0\pada\chi^\da_{0\,a}
+2\ii I(1 + 4I\bar\phi)^{-1}\rho^{\b a}_0\rho^\a_{0\,a}
\partial_{(\alpha\dot\alpha} a^\da_{\b)}\nn \\
&& +\, 2I(1 + 4I\bar \phi)^{-1}(f^{ak}_0f_{0\,ak})\Box \bar\phi
+4\ii I(1 + 4I\bar \phi)^{-1}(\rho_0^{\alpha a}f_{0\,ak})\pada
\bar\psi^{\dot\alpha k}\,. \lb{L12}
\eea
Now it is easy to observe that in the total gauge multiplet-hypermultiplet
Lagrangian $L = L_g + L_1$ the last two terms in \p{L12} can be removed by
the appropriate redefinition of the fields $\varphi$ and $\psi^\alpha_k$ of
the U(1) gauge multiplet
\bea
&& \hat\varphi = (1 + 4I\bar\phi)^2\varphi
- 4I(1 + 4I\bar \phi)^{-1}(f^{ak}_0f_{0\,ak})\,,\nn \\
&& \hat\psi^\alpha_k =(1 + 4I\bar\phi)^2 \psi^\alpha_k -
4 I(1 + 4I\bar \phi)^{-1}(\rho_0^{\alpha a}f_{0\,ak})\,. \lb{REDEF}
\eea
In terms of new fields the total on-shell Lagrangian can be rewritten as
the sum of the free gauge multiplet-hypermultiplet  action and the simple
interaction term
\bea
&& L = L_g+L_1 = L_0 + L_{int}\,, \lb{Vqlagr1} \\
&& L_0 = -\sha \hat\vp\Box\bph + \sfrac12\,\partial_m f_0^{ak}\partial_m
f_{0\,ak} -\sfrac{1}{16}f^\ab f_\ab-\ii\hat\j^\a_k\pada\bj^{\da k} \lb{Vqfree}
\\
&&\qquad \; +\, \sfrac{1}{2}\ii\rho^{\a a}_0\pada\chi^\da_{0\,a}+\sfrac14(
\hat{d}_{kl})^2\,, \nn \\
&& L_{int} = -\sfrac{1}{2}\,I\,\bph(1+2I\bph)\,f^\ab f_\ab
+I\,(1 + 4I\bar\phi)^{-1}\,\rho^{\b a}_0\rho^\a_{0\,a}f_\ab \,,\lb{Vqlagr}
\eea
where $f_\ab=\ii\pa_\ada a_\b^\da+\ii\pa_{\b\da} a_\a^\da=(\si_{mn})_\ab f_{mn}$
and $\hat{d}_{kl}=(1+4I\bph)d_{kl}\,$. Note that the corresponding equations
for the fields $\bph, f_0^{ak}, \bj^{\da k}, \rho^{\a a}_0$ and
$\hat{d}_{kl}$ are free.

In the limit $I=0$ the Lagrangian \p{Vqlagr1} is reduced to a sum of free
Lagrangians of the vector gauge multiplet and hypermultiplet, and so it
represents the Euclidean $N{=}(2,2)$ supersymmetric abelian gauge theory.
Hence in this limit it should exhibit a hidden on-shell $N{=}(1,1)$
supersymmetry which forms $N{=}(2,2)$ supersymmetry together with the
manifest $N{=}(1,1)$. At $I\neq 0$ the Lagrangian \p{Vqlagr1} can be treated
as a QS-deformed version of the $N{=}(2,2)$ gauge theory Lagrangian, and it
is expected to respect half of the original $N{=}(2,2)$ supersymmetry.
Indeed, it can be checked that \p{Vqlagr1} is invariant, up to a total
derivative, under the following extra $N{=}(1,0)$ supersymmetry:
\bea
&&\d_\eta\hat\vp= 0\,,\q\d_\eta\bph=-2 \eta^{\a a}\rho_{0\a a}\,,\q
\d_\eta a_\ada= 2\eta^a_\a\chi_{0a\da}\,,\q\d_\eta\hat\j^\a_k= 0\,,
\q\d_\eta\hat{d}_{kl}= 0\,,\nn\\
&&  \d_\eta \bj^{\da k}= 2\ii\eta^a_\a\pa^\ada
f^k_{0a}\,, \q \d_\eta f^{ak}_0= -2\eta^{\a a}
\hat\j^k_\a\,,\q\d_\eta\chi^\da_{0a}=-2\ii\eta_{\a a}\pa^\ada\hat\vp\,,
 \nn \\
&&\d_\eta \rho^{\a a}_0
=(1+4I\bph)^2\eta^a_\b f^\ab-8I(1+4I\bph)^{-1}\eta_\b^a\rho^{\b b}_0
\rho^\a_{0b}\,,\lb{addsusy}
\eea
where $\eta^{\a a}$ are the corresponding Grassmann parameters. In terms of
the redefined fields the manifest $N{=}(1,0)$ supersymmetry is realized by
the transformations
\bea
&&\d_\eps\hat\vp=2\eps^{\a k}\hat\j_{\a k},\q\d\bph= 0\,,\q
\d_\eps a_\ada=2\eps^k_\a\bj_{\da k}\,,\nn\\
&&\d_\eps\hat\j^\a_k=(1+4I\bph)\eps^{\a l}\hat{d}_{kl}+\sha(1+4I\bph)^2
\eps_{\b k}f^\ab-4I(1+4I\bph)^{-1}\eps_{\b k}(\rho_0^{\a a}\rho^\b_{0 a})\,,
\nn\\
&&\d_\eps \bj^{\da k}=\ii\eps_\a^k\pa^\ada\bph\,,\q
\d_\eps\hat{d}_{kl}=\ii(1+4I\bph)
(\eps^\a_k\pada\bj^\da_l+\eps^\a_l\pada\bj^\da_k)\,,\nn\\
&&\d_\eps f^{ak}_0=\eps^{\a k}\rho_{0\a}^a\,,\q
\d_\eps\chi^\da_{0a}=-2\ii\eps_\a^k\pa^\ada f_{0ak}\,, \q
\d_\eps \rho^{\a a}_0= 0\,.\lb{mansusy}
\eea
The additional $\eta$-transformations commute on-shell with themselves and
also with the  $N{=}(1,0)$ $\eps$-transformations, for instance,
\bea
&&(\d_\eps\d_\eta-\d_\eta\d_\eps) f^{ak}_0=2\eta^{\a a}
(1+4I\bph)\eps_{\a l}\hat{d}^{kl}= 0\,,\;(\d_\eta\d_\eps-
\d_\eps\d_\eta)\hat\j^\a_k=-8I\eta^{\b a}\rho_{0\b a}
\eps^{\a l}\hat{d}_{kl}\nn\\
&&-\,\ii(1+4I\bph)^2\eps_{\b k}[\eta^{\b a}(\pa^\ada\chi_{0\da a}-4\ii\,I
(1+4I\bph)^{-1}\rho_{0\g a}f^{\g\a})+(\a\leftrightarrow\b)]=0\,,\\
&&(\d_{\eta_2}\d_{\eta_1}-\d_{\eta_1}\d_{\eta_2})\bj^{\da k}=
-4\ii\eta^{\b a}_1\eta_{2\b a}\pa^\ada\hat\j^k_\a = 0\,,\nn\\
&&(\d_{\eta_2}\d_{\eta_1}-\d_{\eta_1}\d_{\eta_2})a_\ada=
4\ii\eta^{\b a}_1\eta_{2\b a}\pada\hat\vp\,,
\eea
where we used the equations of motion derived from the Lagrangian
\p{Vqlagr1}. The last bracket yields a composite gauge transformation of
the gauge field.

Presumably, the transformations \p{addsusy} can be derived from a superfield
transformation which is analogous to the one used in the the $N{=}2$
superfield formulation of $N{=}4$ gauge theory in the Minkowski space
\cite{GIOS}. We did not elaborate on this point.

\setcounter{equation}0

\section{Deformed charged hypermultiplet}

According to eqs. \p{GaugeAc}, \p{fundCov}, the superfield action for
the case in question reads
\be
S_e(V,q) = \sfrac12\int\!\diff u\,\diff\zeta^{-4}\ q^+_a
\left(D^{++}q^{+\,a} + \frac{1}{2}\,e [\Vp,q^{+a}]_\star-\frac{1}{2}\,e
(\tau_3)^a_b\{\Vp,q^{+b}\}_\star\right). \lb{SFact2}
\ee

The detailed component structure of this action can be found like in the
previous case, applying the general formulas \p{CommStar}, \p{AntcommStar},
inserting the $\theta^+, \bar\theta^+$ expansions \p{WZ}, \p{qexpan} and
performing integration over the Grassmann and (at the final step) harmonic
variables. Solving the harmonic equations for the auxiliary fields, we finally
express $q^{+a}$ in terms of off-shell fields of the gauged multiplet and
the physical hypermultiplet fields:
\bea
&&q^{+a}_e=u^+_kf^{ak}+\tpa\rho_{\a}^a+(\tp)^2u^-_kg^{ak}
+\btp_\da[\chi^{\da a}+(\tp)^2u^-_ku^-_l\si^{\da akl}] \nn \\
&&+ \, \tp\si_m\btp
r^{ak}_m u^-_k +(\btp)^2[u^-_kh^{ak}+\tpa u^-_ku^-_l\Sigma_\a^{a\,kl}+(\tp)^2u^-_ku^-_lu^-_j
X^{a\,klj}]\,,
\eea
where
\bea
&&g^{ak}= e(\tau_3)^a_b\bph f^{bk},\q r^{ak}_{m}=2\ii(1+2e I\bph)\pa_mf^{ak}
+2e (\tau_3)^a_b A_mf^{bk}\,, \nn\\
&&h^{ak}= -4\ii e IA_m\pa_mf^{ak}+e (\tau_3)^a_b\left(\phi f^{bk}+2eI^2\bph\Box
f^{bk}\right),\nn\\
&&\si^{\da a\,kl}=2e(\tau_3)^a_b \bJ^{\da(k}f^{bl)}\,,\q \Sigma_{\a}^{akl}
=-4\ii e I\bJ^{\da(k}\pada f^{al)}+2e(\tau_3)^a_b\J^{(k}_\a
f^{bl)}\,, \nn \\
&& X^{aklj}= e(\tau_3)^a_b\cD^{(kl}f^{bj)}\,.
\eea

The  deformed residual U(1) gauge transformations of the charged
hypermultiplet components are
\bea
&&\d_r f^{ak}=\ii a\,e (\tau_3)^a_b\,f^{bk}\,,\q\d_r\rho_{\a}^a=\ii a\,e
(\tau_3)^a_b\,\rho_{\a}^b\,,\nn\\
&&\d_r\chi^{\da a}=\ii a\,e (\tau_3)^a_b\,\chi^{\da b}-2e I\pa^\ada a\,
\rho_{\a}^a\,.\lb{ChargeG}
\eea
The corresponding component ubroken $N{=}(1,0)$ supersymmetry transformations
read
\bea
&&\d_\eps f^{ak}=\eps^{\a k}\rho_{\a}^a\,,\q \d_r\rho^a_{\a}=2\eps^k_\a
g^a_{k}=2\eps^k_\a\,e (\tau_3)^a_b\, \bph\,f^b_{k}\,,\nn\\
&&\d_\eps\chi^a_{\da}=-\eps^\a_kr^{ak}_{\ada}=-2\eps^\a_k[\ii(1+2e I\bph)
\pada f^{ak}+ e (\tau_3)^a_bA_\ada f^{bk}]\,. \lb{Chargesusy}
\eea

The SW transform for the charged hypermultiplet fields is given
by the relations which are similar to those for the `adjoint' hypermultiplet,
eqs. \p{SWadj}, though slightly differ in some coefficients
\bea
&&f^{ak}_{0}=(1+2eI\bph)f^{ak}\,,\qq \rho^{\a a}_{0}=(1+2eI\bph)
\rho^{\a a}\,,\nn\\
&&\chi^{a}_{\da 0}=\chi^{a}_{\da}-2eI(1+4eI\bph)^{-1}A_\ada\rho^{\a a}+
4eI(1+4eI\bph)^{-1}\bJ_{\da k}f^{ak}\,.\lb{SWch}
\eea
The gauge and supersymmetry transformations of fields $f^{ak}_{0},\;
\rho^{\a a}_{0}$ and $\chi^{a}_{\da 0}$ look just as the $I=0$ limit of
\p{ChargeG}, \p{Chargesusy}. While checking this, one should take into
account that in the considered case of $e \neq 1$ the gauge and supersymmetry
transformations of the gauge multiplet fields are obtained by
the replacement $I \rightarrow eI$ in eqs. \p{ResComp} and \p{defsusy}.

The deformed charged hypermultiplet Lagrangian for the physical fields
is given by
\bea
&&L_e=\sha(1+4e I\bph)\pa_mf_{ak}\pa_m f^{ak}+ \ii e (\tau_3)_a^bA_mf_{bk}
\pa_mf^{ak}+\sha\,e^2 (A_m)^2(f^{ak})^2 \nn \\
&& +\, \sha\,e^2 \phi\bph (f^{ak})^2
+ I^2e^2 (f^{ak})^2\Box(\bph^2)-\sha\,e (\tau_3)^a_bf^k_{a}f^{bl}\cD_{kl}
+2\ii e I\bJ^\da_k\rho^\a_{a}\pada f^{ak} \nn \\
&& +\, e (\tau_3)^a_b\J^\a_k\rho_{\a a}f^{bk}
+e (\tau_3)^a_bf^k_{a}\bJ_{\da k}\chi^{\da b}+\sfrac{1}{2}\ii(1+2eI\bph)
\rho^{\a a}\pada\chi_{a}^\da \nn \\
&& -\,\sha\,e (\tau_3)^a_b\rho^\a_{a} A_\ada\chi^{\da b}+
\sfrac14\,e (\tau_3)^a_b(\bph\chi_{\da a}\chi^{\da b}
+\phi\rho^\a_a\rho^b_\a)
+\ii e I\rho^{\a a}A_m\pa_m\rho_{\a a}\nn \\
&& +\, \sfrac{1}{2}\ii\,e I\rho^{\b a}\rho^\a_{a}\pa_{(\a\da}
A^\da_{\b)} +\,I^2\,e (\tau_3)^a_b\bph\pada\rho_{\b a}\pa^{\b\da}\rho^{\a b}
 \,. \lb{chargeL}
\eea
It has to be combined with the $e \neq 1$ modification of the U(1) gauge
theory Lagrangian \p{LAGRgauge} rewritten in terms of the original deformed
fields. As distinct from the case of neutral hypermultiplet, passing to the
undeformed fields in the total action with the help of the transformation
\p{SWch} and the $e\neq 1$ version of \p{SWtran} does not give rise to
radical simplifications. Here we present the scalar potential of the model
in terms of the deformed fields. It arises as the result of integrating out
the gauge multiplet auxiliary field ${\cal D}^{kl}$ from the corresponding
piece of the total action
\be
\sfrac14(1+4eI\bph)^{-2}\cD^{kl}\cD_{kl}-\sha\,e(\tau_3)^a_bf^k_{a}f^{bl}
\cD_{kl}+\sha\,e^2\phi\bph(f^{ak})^2\,,\lb{potD}
\ee
and is given by the following positively-definite expression
\bea
&&V=\sfrac18\,e^2(f^{ak})^2[(1+4eI\bph)^2(f^{ak})^2+4\phi\bph]\,.\lb{Potent}
\eea

It is worth noting that there is one more mechanism of generating scalar
potentials in the considered deformed hypermultiplet-gauge multiplet systems.
Namely, one can add to the superfield actions \p{GaugeAc1} or \p{SFact2} the
analytic Fayet-Iliopoulos (FI) superfield term
\be
S_{FI} =\int\!\diff u\,\diff\zeta^{-4}\,\left[3\ii c^{++} + c_0
(\bar\theta^+)^2\right]\Vp \,, \lb{FI}
\ee
where $c^{++} = c^{(ik)}u^+_iu^+_k\,, \widetilde{c^{++}} = c^{++}\,$, and
$c^{(ik)}$ and $c_0$ are some harmonic-independent constants (the numerical
coefficient 3 was introduced for further convenience). This term is
manifestly invariant under $N{=}(1,0)$ supersymmetry and, up to a total
harmonic derivative, under gauge transformations \p{U1V} or \p{U1Vmod}. In
WZ gauge for $\Vp$ FI-term gives rise to the following contribution to the
component Lagrangian
\be
\ii c^{kl}{\cal D}_{kl} + c_0\,\bph\,,
\ee
which, being combined with \p{potD}, results in the SU(2)$_R$-breaking
addition to \p{Potent}
\be
\Delta V = \left[e(\tau_3)^a_b(f^k_af^{bl}c_{kl}) + c^2\right](1 +
4eI\bph)^2 + c_0\,\bph\,.\lb{Pot2}
\ee
It contains the tadpole term $\sim \bph$ which could destabilize the theory
in the quantum case, giving rise to vacuum transitions. Such a term vanishes
under the choice
\be
c_0 = -8\,eI\,c^2\,.
\ee
The same mechanism applies to the case of the adjoint hypermultiplet
considered in the previous Section. The relevant potential is obtained from
\p{Pot2}, replacing $eI \rightarrow I$ and then setting $e=0$. Note that
such a potential breaks the second (hidden) $N{=}(1,0)$ supersymmetry of
this model.

Finally, let us briefly discuss how mass terms for the hypermultiplets
can be introduced in the QS-deformed case. The mechanism of such a generation
is of the Scherk-Schwarz type and it is similar to the one known in the
$N{=}2$ Minkowski case.

Let us consider the constant U(1) analytic potential
\be
B^{\pp}=\bar m(\tp)^2+m(\btp)^2 \,, \lb{Backgr}
\ee
where $m$ and $\bar m$ are independent real constants (they are mutually
conjugated in the Minkowski case). Like in the undeformed case \cite{Z2,IKZ},
$B^{\pp}$ is a background solution of the deformed U(1) gauge theory
equations. The generation of mass terms can be interpreted as a result of
interaction with these constant background `scalar fields' $m$ amd $\bar m$
in \p{Backgr} \cite{IKZ,BK,BBIK}. It is covenient to define new shifted
scalar fields of the gauge U(1) supermultiplet
\be
\phi=m+ \phi^\prime,\qq \bph=\bar m+\bph^\prime\,. \lb{shift}
\ee
The $N{=}(1,0)$ supersymmetry of both gauge field-hypermultiplet models
survives after this shift, although its realization on the component fields
slightly changes, as can be explicitly seen by substituting \p{shift} into
\p{10defq} and \p{Chargesusy}. After making this shift in the Lagrangians
\p{L1}, \p{chargeL} we see that the genuine mass terms appear only for the
charged hypermultiplet and they survive in the commutative limit. Masses of
the left- and right-handed fermions $\rho^\a_a$ and $\chi^\da_a$ are
proprtional to $m$ and $\bar m$ and so are independent. Also, as a specific
feature of the non-anticommutative case, background $\bar m$ induces proper
renormalizations of the kinetic terms. The only impact of this background
on the Lagrangian of the neutral hypermultiplet is such renormalizations of
the kinetic terms.

The new `free' parts of the superfield actions \p{GaugeAc1} and \p{SFact2}
are obtained by substituting there $B^{\pp}$ for $\Vp$, and the shift \p{shift}
corresponds to decomposing
\be
\Vp = \hat{V}^\pp + B^{\pp}\,.
\ee
While the free part of the massless hypermultiplet actions, i.e. \p{freeAc},
is invariant under the standard realization of $N{=}(1,0)$ supersymmetry,
the free actions with mass terms are invariant under modified supersymmetry
transformations which are combinations of the standard $N=(1,0)$
supertranslations and the particular case of U(1) gauge transformations
\p{U1V} (or \p{U1Vmod} for $e\neq 1$) and \p{adj}, \p{fund} (or \p{adjmod},
\p{fundmod}), with
$$
\hat{\L} = -2 \bar m \eps^-\theta^+\,.
$$
E.g., for the charged hypermultiplet the modified $N{=}(1,0)$ transformations
read
\bea
&&\hat\d_\eps q^+=(\eps^{-\a}Q^+_\a-\eps^{+\a}Q^-_\a)q^+ - e\,\hat\L\star
q^+\,. \lb{susymod-q}
\eea
The full off-shell gauge superfield-hypermultiplet action $S_g+S_e(V,q)$
is invariant with respect to \p{susymod-q} and the corresponding modification
of the $N{=}(1,0)$ transformation of $\Vp$
\bea
&&\hat\d_\eps\Vp=(\eps^{-\a}Q^+_\a-\eps^{+\a}Q^-_\a)\Vp+\Dp\hat\L+e\,[\Vp,
\hat\L]_\star\,.
\eea

The modified transformations close on the constant-parameter subgroup of the
corresponding QS-deformed U(1) group, with the appropriate generator being
identified with the central charge of the $N{=}(1,0)$ superalgebra. Since
$\star$-commutators of a constant parameter $\Lambda$ with both $\Vp$ and
$q^{+a}$ are vanishing, the modified $N{=}(1,0)$ transformations have a
non-zero closure only on the charged hypermultiplet and the anticommutator
of the corresponding supercharges is proportional to
$$
\sim (\tau_3)^a_b q^{+b}\,,
$$
precisely as in the undeformed case. The free massive hypermultiplet actions
are invariant just under the modified $N{=}(1,0)$ transformations and by no
means under the original ones (which mix the free actions with the
interaction terms $\sim \hat{V}^\pp $).

It is worth pointing out that this modified $N{=}(1,0)$ supersymmetry is
present in the theory with the deformation operator constructed out of the
standard nilpotent $N{=}(1,0)$ supercharges $Q^\pm_\alpha$ \p{Qgen}.

As already mentioned in Section~3, in the case of the complex hypermultiplet
there are two mutually commuting rigid U(1) groups. In addition to
the  transformation \p{fund4} on the complex
hypermultiplet, one can consider the folowing SU(2)$_{PG}$-breaking
gauge transformation of the same complex superfield:
\be
\d_{\L^\prime}q^{+a}=\sfrac{e}{2}q^{+b}\star\L^\prime[\d^a_b+(\tau_3)^a_b] \,,
\ee
where $\L^\prime$ is an independent analytic superfield parameter.
This `right' transformation commutes with the `left' transformation
\p{fund4}. So we are led to introduce two independent deformed U(1)
gauge potentials, interacting with the complex hypermultiplet superfield,
or e.g. one potential $\Vp$ for the first gauge group and the constant
background potential for the second one. In the second case we shall gain
SU(2)$_{PG}$ breaking mass terms which cannot be generated by shifts of
the scalar fields of the `left' gauge multiplet. Of course we could introduce
background fields for the first U(1) group and gauge the second one, then
we could generate SU(2)$_{PG}$ invariant mass terms and `right' gauge
interaction.

Finally we note that the potential \p{Pot2} induced by the FI-term provides
another independent $N{=}(1,0)$ supersymmetric source of generating masses
for the scalar physical fields of the hypermultiplet (and simultaneously for
the field $\bph$). The superfield FI-term \p{FI} is invariant under both the
original and central-charge modified $N{=}(1,0)$ supersymmetries.

\section{Conclusions}

We have constructed, for the first time, the $N{=}(1,1)$ non-anticommutative 
singlet Q-deformation (QS-deformation) of a hypermultiplet coupled to a U(1) 
gauge supermultiplet in Euclidean~$\mR^4$. We found that there exist two 
essentially different possibilities of implementing the QS-deformed U(1) gauge 
action on the hypermultiplet, namely the `adjoint' and the `fundamental' 
representation. These correspond, respectively, to QS-deformations of systems 
with neutral and charged hypermultiplets. For both options we constructed 
superfield and component actions, explicitly gave the transformations of 
unbroken $N{=}(1,0)$ supersymmetry and presented the appropriate Seiberg-Witten
transformations to the undeformed fields. The first system was shown to possess
an additional (hidden) $N{=}(1,0)$ supersymmetry, thus providing a 
QS-deformation of Euclidean gauge theory with $N{=}(2,2)$ supersymmetry. 
We then studied the effect of adding Fayet-Iliopoulos terms to the 
gauge-superfield-hypermultiplet actions constructed and presented the relevant
scalar potentials. We also demonstrated the possibility of generating a mass 
term for the charged hypermultiplet via the Scherk-Schwarz mechanism related 
to the appearance of a central charge in the $N{=}(1,0)$ superalgebra. 
This central charge is identified with the generator of one of the global U(1)
symmetries realized on the hypermultiplet.

Our results can be extended along several directions. In particular, 
it would be interesting to detect possible phenomenological uses of the 
considered QS-deformations as well as of their non-abelian generalizations. 
Deformations of this sort may possibly provide a geometrical way of introducing
soft supersymmetry breaking. Of course, an appropriate Wick rotation has to
be performed to connect with realistic models in Minkowski space.
A closely related quantum issue are the renormalization and finiteness 
properties of our coupled gauge multiplet-hypermultiplet systems, which may
be investigated as in the $N{=}(\sha,\sha)$ case \cite{Rev2}. The $N{=}(2,2)$ 
gauge theory is the Euclidean analog of $N{=}4$ super Yang-Mills 
which supplied the first example of an ultraviolet-finite quantum field theory
and displays various other remarkable properties (e.g.~in the stringy context
of the AdS/CFT correspondence). It is of obvious interest to study the 
$N{=}(2,2)$ gauge theory from these angles and to examine whether its nilpotent
deformation as presented in Section~3 and its non-abelian generalization
preserve the basic quantum and geometric properties of the undeformed theory. 
We recall that nilpotent deformations do not induce noncommutativity for the 
bosonic spacetime coordinates; hence, such deformed quantum theories are 
expected not to suffer from the typical non-locality problems such as UV-IR 
mixing. The issues of their vacuum structure and classical solutions, 
their non-abelian generalization and the relationship between their Coulomb 
and Higgs branches all seem to be interesting tasks for future work. 
           
As another possible application of the deformed minimal coupling of a U(1) 
gauge multiplet with hypermultiplets explicitly worked out in this paper, 
we mention a generalization of the quotient approach to constructing 
hyper-K\"ahler metrics in $N{=}2$ supersymmetric sigma models 
(see e.g.~\cite{HK}).
Its basic ingredient is a non-propagating $N{=}2$ U(1) gauge multiplet 
coupled to hypermultiplets. Using the relations given in Sections~3 and~4 
it is easy to generalize this approach to QS-deformed $N{=}(1,1)$ theories 
and to explore the possible impact on the hyper-K\"ahler target space geometry.
Finally, we point out that it is also desirable to elaborate on the 
implications of more general (non-singlet) Q-deformations in $N{=}(1,1)$ 
(and perhaps $N{=}(2,2)$) supersymmetric systems with hypermultiplets.

\section*{Acknowledgements}

This work was partially supported
by the INTAS grant 00-00254,
by the DFG grants 436 RUS 113/669-2 and Le-838/7,
by the RFBR grants 03-02-17440 and 04-02-04002,
by the NATO grant PST.GLG.980302 and
by a grant of the Heisenberg-Landau program. E.I. and B.Z. thank
the Institute of Theoretical Physics of the University of Hannover for
the warm hospitality at the initial stages of this study.

%\vfill\eject


\begin{thebibliography}{99}
\addtolength{\itemsep}{-5pt}

\bibitem{SW} N. Seiberg, E. Witten, 
JHEP {\bf 9909} (1999) 032 {[\tt hep-th/9908142]}. 
\bibitem{Rev1} M.R. Douglas, N.A. Nekrasov,  
Rev. Mod. Phys. {\bf 73} (2001) 977 {[\tt hep-th/0106048]}.
\bibitem{OVa} H. Ooguri, C. Vafa,
%{\sl The C-deformation of gluino and non-planar diagrams},
Adv. Theor. Math. Phys. {\bf 7} (2003) 53 {\tt [hep-th/0302109]}.
\bibitem{BGN} J. de Boer, P.A. Grassi, P. van Nieuwenhuizen,\\
%{\sl Noncommutative superspace from string theory}
Phys. Lett. {\bf B 574} (2003) 98 {\tt [hep-th/0302078]}.
\bibitem{Se} N. Seiberg,
JHEP {\bf 0306} (2003) 010 {\tt [hep-th/0305248]}.
\bibitem{BS} N. Berkovits, N. Seiberg, 
JHEP {\bf 0307} (2003) 010 {[\tt hep-th/0306226]}.
\bibitem{Rev2} M.T. Grisaru, S. Penati, A. Romagnoni,\\ 
Class. Quant. Grav. {\bf 21} (2004) S1391 {[\tt hep-th/0401174]}. 
\bibitem{FL} S. Ferrara, M.A. Lled\'o, 
JHEP {\bf 05} (2000) 008 {\tt [hep-th/0002084]}.
\bibitem{KPT} D. Klemm, S. Penati, L. Tamassia,\\
Class. Quant. Grav. {\bf 20} (2003) 2905 {\tt [hep-th/0104190]}.
\bibitem{FLM} S. Ferrara, M.A. Lled\'o, O. Maci\'a,
JHEP {\bf 09} (2003) 068 {\tt [hep-th/0307039]}.
\bibitem{ILZ} E. Ivanov, O. Lechtenfeld, B. Zupnik,  
JHEP {\bf 0402} (2004) 012 {\tt [hep-th/0308012]}.
\bibitem{FS} S. Ferrara, E. Sokatchev,  
Phys. Lett. {\bf B 579} (2004) 226 {\tt [hep-th/0308021]}.
\bibitem{FILSZ} 
S. Ferrara, E. Ivanov, O. Lechtenfeld, E. Sokatchev, B. Zupnik,\\
{\sl Non-anticommutative chiral singlet deformation of N=(1,1) gauge theory},\\
%CERN-PH-TH/2004-032, ITP-UH-10/04, LAPTH-1041/04 
Nucl. Phys. {\bf B} in press {\tt [hep-th/0405049]}.
\bibitem{AIO} T. Araki, K. Ito, A. Ohtsuka, 
JHEP {\bf 0401} (2004) 046 {[\tt hep-th/0401012]}.
\bibitem{AI} T. Araki, K. Ito, 
%{\sl Singlet deformation and non(anti)commutative N=2 supersymmetric
%U(1) gauge theory},
Phys. Lett. {\bf B 595} (2004) 513 {\tt [hep-th/0404250]}.
\bibitem{KS} S.V. Ketov, S. Sasaki, 
%{\sl BPS-type equations in the
%non-anticommutative N=2 supersymmetric U(1) gauge theory.}
Phys. Lett. {\bf B 595} (2004) 530 {\tt [hep-th/0404119]}; \\
Phys. Lett. {\bf B 597} (2004) 105 {\tt [hep-th/0405278]}; \\
{\sl SU(2)${\times}$U(1) non-anticommutative N=2 supersymmetric gauge theory}\\
{\tt [hep-th/0407211]}.
\bibitem{BFPL} M. Billo, M. Frau, I. Pesando, A. Lerda, 
JHEP {\bf 0405} (2004) 023 {[\tt hep-th/0402160]}.
\bibitem{GIK1} A. Galperin, E. Ivanov, V. Ogievetsky, E. Sokatchev, \\
JETP Lett. {\bf 40} (1984) 912 [Pis'ma ZhETF {\bf 40} (1984) 155]; \\
A. Galperin, E. Ivanov, S. Kalitzin, V. Ogievetsky, E. Sokatchev,\\
Class. Quant. Grav. {\bf 1} (1984) 469.
\bibitem{GIOS} A. Galperin, E. Ivanov, V. Ogievetsky, E. Sokatchev,\\
{\sl Harmonic superspace}, Cambridge University Press, 2001, 306 p.
\bibitem{IZ} E.A. Ivanov, B.M. Zupnik,\\
{\sl Non-anticommutative deformations of N=(1,1) supersymmetric theories},\\
talk at ``Classical and quantum integrable systems'', 
Dubna, 26-29 January, 2004 \\
{\tt [hep-th/0405185]}.
\bibitem{SWo}C. S\"amann, M. Wolf, 
%{\sl Constraint and super Yang-Mills equations
%on the deformed superspace $\mR_\hbar^{(4|16)}$}, Preprint UTP-UH-05/04
JHEP {\bf 0403} (2004) 048 {\tt [hep-th/0401147]}.
\bibitem{Z2} B.M. Zupnik, 
Sov. J. Nucl. Phys. {\bf 44} (1986) 512 [Yad. Fiz. {\bf 44} (1986) 794].
\bibitem{IKZ} E. Ivanov, S. Ketov, B. Zupnik,  
Nucl. Phys. {\bf B 509} (1998) 53 {\tt  [hep-th/9706078]}.
\bibitem{BK} I.L. Buchbinder, S.M. Kuzenko, \\
Class. Quant. Grav. {\bf 14} (1997) L157 {\tt [hep-th/9704002]}.
\bibitem{BBIK} E.I. Buchbinder, I.L. Buchbinder, E.A. Ivanov, S.M. Kuzenko,\\
Mod. Phys. Lett. {\bf A13} (1998) 1071 {\tt [hep-th/9803176]}.
\bibitem{HK} N.J. Hitchin, A. Karlhede, U. Lindstr\"om, M. Ro\v{c}ek, \\
Commun. Math. Phys. {\bf 108} (1987) 535; \\
A. Galperin, E. Ivanov, V. Ogievetsky, P.K. Townsend, \\
Class. Quant. Grav. {\bf 3} (1986) 625.

\end{thebibliography}
\end{document}